\documentclass[preprint,aps,nofootinbib,amssymb,amstex,amsfonts]{revtex4}
\newcommand{\ket}[1]{|#1\rangle}                	%
\newcommand{\brac}[1]{\langle #1|}              	%
\newcommand{\bra}[1]{\langle #1}               		%
\newcommand{\grkbf}[1]{\mbox{\boldmath $#1$}}        	%
\newcommand{\pd}{\partial}                           	%
\newcommand{\nx}[1]{\mathbf{#1}}                 	%
\newcommand{\hil}{{\cal H}}              		%
\newcommand{\hilN}{{\cal H}_{N}}             		%
\newcommand{\hilNN}{{\cal H}_{N^2}}          		%
\newcommand{\De}{\text{\textbf{\textsf{D}}}_{\epsilon}} 	
\newcommand{\Tqp}{\hat{T}_{(q,p)}}           		%
\newcommand{\hrho}{\hat{\rho}}               		%
\newcommand{\hA}{\hat{A}}                		%
\newcommand{\hB}{\hat{B}}                		%
\newcommand{\hM}{\hat{M}}                		%
\newcommand{\hR}{\hat{R}}				%
\newcommand{\hL}{\hat{L}}				%
\newcommand{\hS}{\hat{S}}				%
\newcommand{\hF}{\hat{F}}				%
\newcommand{\SOp}[1]{\text{\textbf{\textsf{#1}}}}       
\newcommand{\II}{\hat{I}}        			
\newcommand{\TT}{\mathbb{T}^{2}}        		%
\newcommand{\RR}{\mathbb{R}^2}          		%
\newcommand{\s}{\text{\textbf{\textsf{S}}}}		%
\newcommand{\Ldos}{\mathbb{L}^2}        		%
\newcommand{\hT}{\hat{T}}               		%
\newcommand{\hP}{\hat{P}}               		%
\newcommand{\hU}{\hat{U}}       			%
\newcommand{\hV}{\hat{V}}       			%
\newcommand{\hQ}{\hat{Q}}       			%
\newcommand{\U}{\hat{{\cal U}}}     			%
\newcommand{\ie}{\mbox{{\em i.e.\/}} }          	%
\newcommand{\etal}{\mbox{{\em et al.\/}} }      	%
\newcommand{\tr}{\mbox{Tr}}                 		%
\newcommand{\equa}[1]{Eq.~(\ref{#1})}       		%
\newcommand{\fig}[1]{Fig.~\ref{#1}}         		%
\newcommand{\sect}[1]{Sec.~\ref{#1}}			%
\newcommand{\rmi}{{\rm i}}
		
\usepackage{graphicx,graphics,rotating}     	
\usepackage{dcolumn}                    		
\usepackage{bm}                		
\usepackage[german,english]{babel}			%
\begin{document}
\title{SPECTRAL APPROACH TO CHAOS AND QUANTUM-CLASSICAL CORRESPONDENCE IN
QUANTUM MAPS}

\author{Ignacio Garc\'\i a-Mata}
\email[Email address: ]{garciama@tandar.cnea.gov.ar}
\affiliation{%
Departamento de F\'{\i}sica, Comisi\'{o}n Nacional de Energ\'{\i}a At\'{o}mica.
Avenida del Libertador 8250 (C1429BNP), Buenos Aires, Argentina.
}%
\author{Marcos Saraceno}
\affiliation{%
Departamento de F\'{\i}sica, Comisi\'{o}n Nacional de Energ\'{\i}a At\'{o}mica.
Avenida del Libertador 8250 (C1429BNP), Buenos Aires, Argentina.
}%
\affiliation{%
Escuela de Ciencia y Tecnolog\'\i a, Universidad Nacional de San
Mart\'\i n. Alem 3901 (B1653HIM), Villa Ballester, 
Argentina.}
\date{\today}%

\begin{abstract}
Correspondence in quantum chaotic systems is lost in short time scales. 
Introducing  some noise we study the spectrum of the resulting coarse grained propagaor
of density matrices. Some differen methods to compute the spectrum are reviewed. 
Moreover, the relationship between the eigenvalues of the coarse-grained superoperator 
and the classical Ruelle-Pollicott resonances is remarked. As a concequence, 
classical decay rates in quantum time dependent quantities appear. 
\end{abstract}

\maketitle


\section{Introduction}
Since the beginings of quantum theory, the problem of the transition 
from the quantum to the classical world has posed interesting and challenging questions.
The concept of decoherence\cite{zurekRMP} has introduced some clarity, 
yet many questions remain to be answered.

Quantum chaos studies the emergence of classical chaotic behavior in quantum
systems as $\hbar\to 0$.
The introduction of Gutzwiller's trace formula\cite{gutz} and the developement 
of periodic orbit theories\cite{chaosbook,vergini,heller} 
and in addition the understanding of the influence of random matrix theories\cite{bohigas}
gave quantum chaos a boost of interest. 
One key feature of quantum chaotic systems is that correspondence 
is lost in very short time scales (Eherenfest time\cite{berman}). 
The introduction of noise (in a controlled way) restores correspondence 
and allows to study it from the point of view of the appearance 
of classical properties from the quantum systems. 
In this way, the classical 
Lyapunov exponent plays an important role in the medium time decay of 
time dependent quantitites like the linear entropy\cite{bianucci,garma} 
and the Loschmidt echo\cite{jalabert,losch}. 

The study of this correspondence for long times, like for example the asymptotic
behavior of correlation functions, requires the understanding of the spectral
properties of both quantum and classical propagators and their relationship in
the $\hbar\to 0$ limit.

In this review the latter approach is discussed in detail although the former is 
also addressed. 
Some recent techniques to compute the spectrum of both the 
classical and the quantum propagator of densities are described. 
We describe in detail some efficient methods used by the authors 
recently\cite{garma,garma2,aolita}.
The systems are quantum maps on the torus.  
By introducing diffusion in the form of Gaussian noise we obtain a 
{\em coarse grained\/} quantum propagator\cite{nonn}. 
The Gaussian noise introduces a natural truncation that 
is understood using phase space representation theory 
(chord funtion and Wigner function). In addition, we develop an iteration method that 
efficiently provides the leading part of the spectrum. Relation to the classical 
Ruelle-Pollicott resonances is then established by taking the limit $\hbar\to 0$ 
and also by studying the long time decay rates of several quantities.

This review is organized as follows. In \sect{sect:phasespace} we define the phase space 
we will work with and summarize some operator theoretic properties of representations 
in phase space. A brief introduction to the theory of Open Quantum systems is provided 
in \sect{sec:OQD}. 
In sections \ref{sec:qmaps} and \ref{sec:Oqmaps} 
quantum maps and the model used for open quantum maps are described.
The kind of Gaussian noise used, in term of phase space translations and how it acts
when composed with a unitary superoperator, 
yielding a coarse grained propagator, is developed in \sect{sec:diff} 
while the issue of correspondence, specifically through the relations 
between classical and quantum spectrum is addressed in \sect{sec:ch-corr}. 
Finally in \sect{sec:num}, some numerical methods to compute the 
spectrum are described. 
The consequences for the asymptotic decay for long times are discussed in \sect{sec:LT}.
\section{Phase space}
	\label{sect:phasespace}
\subsection{Torus quantization}
All the calculations are made for systems 
with a bounded phase space, which for simplicity we take to be a 2-Torus ($\TT$).

There are some well-known features that appear in the quantization of
$\TT$.
Since the area is finite, and normalized to unity, 
the Hilbert $\hilN$ space is finite and of dimension $N$.
The value of $N$ determines an effective Planck constant 
\begin{equation}
\hbar=\frac{1}{2\pi N}.
\end{equation}
The position and momentum bases are discrete sets
\begin{equation}
\begin{array}{c}
{\cal B}_q=\{\ket{q_i},q_i=0,1,\ldots N-1\}\\
{\cal B}_p=\{\ket{p_j},p_j=0,1,\ldots N-1\}
\end{array}
,
\end{equation}
which define an $N\times N$ grid $G_N$. We identify with $\ket{q_i}$ the position  
$q_i/N$ and with $\ket{p_j}$ the momentum $p_j/N$.
They are related by the discrete Fourier transform (DFT) of dimension $N$,
\begin{equation}
\bra{p}\ket{q}=\frac{1}{\sqrt{N}}\,e^{-(2\pi \rmi/N)qp}.
\end{equation}

Pure states are represented by $N$-dimensional vectors $\ket{\phi}$ in $\hilN$.
Mixed states are best described by the density operator $\hrho$. 
If we identify Liouville space $\hilNN$ with the space of comlplex $N\times N$
matrices, then density operators associated to states in $\hilN$ form a subset of $\hilNN$
of self-adjoint, positive semidefinite matrices, with unit trace. Moreover  
\begin{equation}
\tr(\hrho^2)\leq 1,
\end{equation}
with the equality holding iff the state is pure.

We now define the translation operators on $\TT$. In $\RR$ the Weyl  group
defines the translation operator of $q_1$ in position and $p_1$ in momentum,
as
\begin{equation}
\Tqp=e^{-(\rmi /\hbar)(q_1\hP-p_1\hQ)}.
\end{equation}

Translations on the torus are somewhat harder to define because infinitesimal
operators $\hQ$ and $\hP$ with the canonical commutation rules cannot be
defined in a discrete Hilbert space. 
Nevertheless, finite cyclic shifts $\hU$ and $\hV$ such that, 
if $\ket{q}$ and $\ket{p}$ are 
position and momentum eigenstates respectively, can be defined\cite{schwinger} as
\begin{equation}
\begin{array}{ccccccr}
\hU^{q_1}\ket{q}&=&\ket{q+q_1};&\ \ &\hU^{q_1}\ket{p}&=&e^{-(2\pi\rmi/N)q_1 p}\ket{p};\\
\hV^{p_1}\ket{p}&=&\ket{p+p_1};& &\hV^{p_1}\ket{q}&=&e^{(2\pi\rmi/N)p_1 q}\ket{q} 
\end{array}
.
\end{equation}
These operators satisfy the comutation rule
\begin{eqnarray}
	\label{eq:commute}
\hV^p\hU^q=\hU^q\hV^p e^{(2\pi \rmi/N)qp}.
\end{eqnarray}
Symmetrizing \equa{eq:commute} one gets the transaltion
operators on the quantized torus\footnote{We do not distinguish the notation
of translations on $\RR$ and $\TT$ because we mainly use the latter.} as a
composition of position and momentum shifts,
\begin{eqnarray}
	\label{eq:trans}
\Tqp&=&\hU^q\hV^p e^{(\rmi\pi/N)qp}\nonumber \\
    &=&\hV^p\hU^q e^{-(\rmi\pi/N)qp}
\end{eqnarray}
with $q,p=0,\ldots,N-1$.
There are $N^2$ operators $\Tqp$ and have the following property
\begin{equation}
\tr(\hT_\alpha^\dag\hT_\beta)=N\delta_{\alpha\beta},
\end{equation}
for all $\alpha, \beta$ (when convenient we use greek letters to rperesent a phase space point, 
{\em e. g.\/} $\alpha\equiv(q,p)$). Therefore,
they constitute a complete orthogonal set in the Hilbert-Schmidt
inner product. Moreover, they have the group composition rule 
\begin{equation}
	\label{eq:composition}
\hT_\alpha\hT_\beta=e^{(\rmi\pi/N)\alpha\wedge\beta}\hT_{\alpha+\beta}
\end{equation}
where $\alpha\wedge\beta$ is the usual wedge product.
\subsection{Phase space representations of states and operators}
	\label{sec:opertor}
The formulation of quantum mechanics in phase space has proven useful
in the understanding of correspondence because phase-space is the natural
stage for classical mechanics. If one looks for a quantum analog of a
probability distribution in phase space the Wigner function appears to be
the best choice. If $\hrho$  is a density operator in a continuous Hilbert
space the Wigner function is usually defined as
\begin{equation}
{\cal W}(q,p)=\int_{-\infty}^{\infty}\frac{dx}{2\pi\hbar} e^{\rmi xp/\hbar}
\brac{q-x/2}\hrho\ket{q+x/2}.
\end{equation}
Although it is not necesarily positive everywhere, this function is real
valued and the integral along any line in phase space yields the correct
marginal distributions.\footnote{Measuring along different lines is the basis of a
widespread technique such as quantum state (and process) tomography.} 
Bertrand and Bertrand \cite{bertrand} showed that
this properties toghether with
\begin{equation}
\tr(\hrho_1\hrho_2)=2\pi\hbar\int{\cal W}_1{\cal W}_2 \,d\alpha,
\end{equation}
uniquely determine the Wigner function.

Quantum discrete phase space descriptions can also be made.
There are many works on different subjects that take advantage of this line
of research.\footnote{See \cite{schwinger,hannay,ozorivas,wooters,Miquel,nature},
to list just a few.} We briefly review general operator theoretic notions of
\emph{representations}.

Operators in $\hilNN$ can be represented by their c-number function in some operator basis.
There are $N^2$ linearly independent operators in $\hilNN$ with the Hilbert-Schmidt inner
product. Let $\{\hS_\alpha\}_{\alpha=0}^{N^2-1}$ be a complete set of operators such that
\begin{equation}
\tr(\hS^\dag_\alpha\hS_\beta)=\delta_{\alpha\beta}
\end{equation}
then an arbitrary operator $\hB$ can be expanded as
\begin{equation}
\hB=\sum_{\alpha=0}^{N^2-1}{\cal S}_B(\alpha)\hS_\alpha.
\end{equation} 
The coeficients of the expansion are
\begin{equation}
{\cal S}_B(\alpha)=\tr(\hS_\alpha^\dag\hB)
\end{equation} 
and ${\cal S}_B$ 
is called the {\em symbol\/} of $\hB$ in the $\hS_\alpha$ 
basis representation. 

If the basis operators are the translations $\hT_\alpha$ then we have
the {\em characteristic function\/} representation
\begin{equation}
\hB=\frac{1}{N}\sum_\alpha {\cal C}_B(\alpha)\hT_\alpha,
\end{equation}
with the symbol
\begin{equation}
	\label{eq:chordfunct}
{\cal C}_B(\alpha)=\tr(\hT_\alpha^\dag\hB),
\end{equation}
The function ${\cal C}_B(\alpha)$ 
is the so-called {\em chord function\/}\cite{ozorivas} of $\hB$, and it owes the name
to its particular geometrical features. 

The translations defined in \equa{eq:trans} are not periodic in the $N\times N$
grid, although they form an orthonormal set. 
If we extend the grid $G_N$  to a $2N\times  2N$ grid $G_{2N}$ then the DFT
of the translations in this grid 
\begin{equation}
\hA_{\beta}=\frac{1}{(2N)^2}\sum_{\alpha=0}^{(2N)^2-1}\hT_\alpha
\,e^{-(2\pi\rmi/2N)\alpha\wedge\beta}
\end{equation}
defines the discrete phase-space point operators. 
There are $4N^2$ of these operators. 
However it can be shown\cite{Miquel} that only
$N^2$ are linearly independent.
The  expansion
\begin{equation}
	\label{eq:weyl}
\hB=N\sum_{\alpha\in G_{2N}}{\cal W}_B(\alpha)\hA_\alpha ,
\end{equation}
determines the discrete Weyl (or
{\em center\/}) representation of $\hB$. The symbol ${\cal W}_\rho$ 
of a density
operator $\hrho$ 
in the Weyl representation is the discrete Wigner function (DWF).
Miquel,
\etal\cite{Miquel} show that the DWF defined 
by \equa{eq:weyl}
has all the desired
properties analogous to those of the continuous Wigner function.
\section{Open Quantum Dynamics}
	\label{sec:OQD}
When studying realistic models of quantum systems one usually thinks in terms
of the specific system of interest (labeled $s$) and a reservoir
(labeled $r$)
with which the system interacts. Their evolution as a whole is unitary and
 (we assume) determined by a Hamiltonian like
\begin{equation}
\hat{H}=\hat{H}_s+\hat{H}_r+\hat{H}_{sr}
\end{equation}
where  $\hat{H}_s$ and $\hat{H}_r$ describe the evolution of the system and the reservoir
independently and $\hat{H}_{sr}$ is the interaction Hamiltonian.
The state of the system is given by the density operator  
$\hrho_{sr}\in\hil_s\otimes\hil_r$.

Standard treatment of quantum systems in the presence of noise, generally in
the form of a reservoir or bath of harmonic oscillators, involves tracing out 
from the unitary evolution of the whole (system plus environment) 
only the reservoir degrees
of freedon yielding a nonlinear partial differential equation or master
equation. When the bath is assumed to fulfill de Born-Markov approximation,
which requires that the relaxation times of the bath are much shorter that
any characteristic time of the evolution of the system, the master equation
can be put in the form
\begin{equation}
	\label{eq:master}
\frac{\partial\hrho_s}{\pd t}=
-\frac{i}{\hbar}[\hat{H},\rho]+\frac{1}{2}\sum_{ij}c_{ij}
\left\{[\hF_i\hrho_s,\hF_j^\dag]+[\hF_i,\rho \hF_j^\dag]\right\},
\end{equation}
where $\hrho_s$ is the density matrix of the system and $\hat{F}_i$ are
system operators. 
When the matrix of coeficients $c_{ij}$ (GKS matrix \cite{GKS}) can be diagonalized 
\equa{eq:master} takes
a simpler, so-called Lindblad\cite{L}, form.
Gorini, \etal\cite{GKS} show that equations of the form
\equa{eq:master} generate a one 
parameter dynamical semigroup of trace-preserving, completely positive quantum
operations. Complete positivity means that a trivial extension of the quantum
operation to a larger Hilbert space yields a positive operation.\footnote{See
Preskill\cite{preskill} or Nielsen and Chuang\cite{chuang}}

A trace preserving, completely positive quantum operation $\s$ can always be
written in a generalized {\em operator sum\/} or Kraus \cite{Kraus} form
\begin{equation}
	\label{eq:kraus}
\s(\rho)=\sum_\mu c(\mu)\hM_\mu\rho \hM_\mu^\dag
\end{equation}
where $c(\mu)\geq 0$ and  $\hM_\mu$ are system operators. If $N$ is the dimension of the state
space, then $\mu$ is no bigger than $N^2$. The map can be shown to be TP if
\begin{equation}
	\label{eq:TP}
\sum_\mu c(\mu)\hM_\mu^\dag \hM_\mu=\II.
\end{equation}
\section{Quantized Maps}
	\label{sec:qmaps}
We will focus our attention on {\em discrete time\/} systems, \ie maps. Instead of modelling
interaction with the environment by proposing the system operators 
$\hat{F}_i$
that appear in the master equation, we directly model the effect the noise has
on the system by giving the noise superoperator specifically in Kraus operator
sum form.

General dynamical  are defined by a first order differential equation
\begin{equation}
\dot{\SOp{x}}=f(\SOp{x})
\end{equation}
where $\SOp{x}$ is a point in phase space. Usually the solution (and the
visualisation) is simplified by means of a discrete map which arise by
observation of the map at fixed time intervals (stroboscopic map) or when the
flow passes transversely through some hypersurface ${\cal P}$
in phase space
(Poincar\'e section). So now time is discrete, and in general labeled $n$,
and the equation is a recurrence, 
\begin{equation}
\SOp{y}_{n+1}=M(\SOp{y}_n), \ \ \ \SOp{y}\in{\cal P}.
\end{equation}

Area preserving maps are characterized by a finite canonical
transformation. The unitary representation of this canonical
transformation \cite{canonical} is what is usually considered the
quantum version of the map. Although there are not general procedures to
obtain this representation, in the real two-dimensional plane $\RR$, 
the procedure is relatively standard.\cite{OdAbook} An aproximation of
the unitary propagator in the semiclassical limit can be written as
\begin{equation}
U(q_1,q_2)=\left(\frac{i}{\hbar}\,\frac{\pd^2 S}{\pd q_1\pd
q_2}\right)^{1/2}\,\exp\left[\frac{i}{\hbar}\,S(q_2,q_1)\right],
\end{equation}
where $S$ is the action along the unique classical path from $q_1$
to $q_2$, and where  for simplicity  we do not consider the
existence of multiple branches and Maslov indices. Only for linear
symplectic maps on $\RR$ this unitary propagator is exact, and
then
 $S$ is minus
the quadratic generating function of the linear transformation.
By different {\em ad-hoc\/} procedures, some of the best known classical
maps have been quantized. Some integrable, such as
translations\cite{schwinger} and shears,
and also chaotic, such as cat maps,\cite{hannay,OdA} baker 
maps,\cite{balazs,saraceno} and the
standard map.\cite{izra} Moreover, noncommuting nonlinear shears, or
``kicked'' maps, can be quantized as well as time dependent periodic
Hamiltonians.\cite{BerryVoros}

On $\TT$ a quantized map is a unitary operator $U\in\hilNN$. 
The evolution of density matrices by this map is given by
\begin{equation}
	\label{eq:densunit}
\hrho_{n+1}=\hU\hrho_{n}\hU^\dag.
\end{equation} 
The last equation can be written as a unitary map $\SOp{U}:\hilNN\to\hilNN$
in Liouville space, that
takes density matrices to density matrices
\begin{eqnarray}
	\label{eq:densunitII}
\hrho_{n+1}&=&\left(\hU\odot\hU^\dag\right)\hrho_{n}\nonumber \\
	   &=&\SOp{U}\hrho_{n}.
\end{eqnarray}
We introduce the $\odot$ symbol to represent {\em adjoint\/} or 
{\em left-right\/} action, 
\begin{equation}
\left(\hU\odot\hU^\dag\right)\hrho\stackrel{\text{def}}{=} Ad(U)\hrho
=\hU\hrho\hU^\dag
\end{equation}
Maps of density matrices to density matrices are called {\em
superoperators\/} or {\em quantum
operations\/}.\footnote{We use indistinctly both denominations.} Further
properties of more general (\ie non-unitary) superoperators are given in the
subsequent sections.
\section{Open Quantum Maps}
	\label{sec:Oqmaps}
We propose as a simplification 
to  model discrete time open quantum evolution 
with a nonunitary propagator (superoperator) $\s$ that consists of
{\em two steps\/}. First the unitary $\SOp{U}$ 
step corresponding to noiseless evolution. Then the
noise acts through a nonunitary superoperator $\De$, where $\epsilon$
characterizes the coupling strength. Thus the evolution given by the two-step
noisy propagator looks like
\begin{equation}
	\label{eq:2step}
\hrho_{n+1}=\s_\epsilon(\hrho_n)=\De\circ\SOp{U}(\hrho_n)
\end{equation}
This type of two-stage schemes have been studied recently in several
works.\cite{garma,garma2,aolita,voros,braun,nonn} They appear naturally in situations
in which the noise is negligible during the unitary evolution. 
Suppose the solution comes from the direct exponentiation of the Lindblad equation. 
Then if to leading order, the unitary part and the noise part commute, 
then the two parts can be separated into a product of two propagators.
One clear physical example is the micromaser\cite{abu}.
There are some other cases like a 
 kicked map in which the interaction takes place between kicks. The case
of a billiard inside a bath, if the interaction with the walls is cosidered
essentially unitary while free propagation is noisy, 
constitutes another example. Another example could be an iterated quantum algorithm,
if the errors due to imperfect gates can be neglected but the channel is noisy.
\section{Diffusive Noise and Coarse-Grained Quantum Propagator}
	\label{sec:diff}
We now proceed to describe the type of noise $\SOp{D}$ considered for the two step
propagator of equation \equa{eq:2step}.	
We will use noise models given explicitly in Kraus form
\begin{equation}
	\label{eq:RUP}
\SOp{D}=\sum_i p_i \U_i\odot\U_i^\dag
\end{equation}
with $\U$ a unitary operator and $p_i\ge 0$. 
From \equa{eq:TP} we need
\begin{equation}
\sum_i p_i=1
\end{equation}
for trace preservation. This condition also implies that the sum in
\equa{eq:RUP} is a convex sum of unitary operators. Any map that can be written
as a convex sum of unitary operators is a {\em random unitary process\/} (RUP).
The physical interpretation of a RUP is simple: {\em the unitary operation
$\U_i\hrho\U_i^\dag$ is applied to the state $\hrho$ with probability $p_i$},
and the sum in\equa{eq:RUP} provides a simple model for a quantum
Langevin-type process.
It is easy to show that a trace preserving RUP is also {\em unital\/} or
identity preserving, that is
\begin{equation}
\SOp{D}(\II)=\II
\end{equation}
which means that the eigenvalue 1 has the maximally mixed state
\begin{equation}
	\label{eq:uniform}
\hrho_\infty=\II/N
\end{equation}
as eigenstate.
Maps that are both TP and unital are called {\em bi-stochastic\/} and they are
{\em contractive maps\/} which means that the spectrum 
contained in the unit circle.

We propose as diffusive noise the following RUP which is diagonal in the chord representation
\begin{equation}
\De={1\over N}\sum_{\alpha}{\cal C}_\epsilon(\alpha)
\hT_\alpha\odot\hT_\alpha^\dag
\end{equation} 
with ${\cal C}_\epsilon(\alpha)$ a Gaussian of width proportional to
$\epsilon$. The chord function of $\De$ is 
\begin{eqnarray}
\tr[\hT_\beta^\dag\De\hT_\beta']&=&
\tr\left[{1\over N}\hT_\beta^\dag\sum_\alpha{\cal
C}_\epsilon(\alpha)\hT_\alpha\hT_\beta'\hT_\alpha^\dag\right]\nonumber\\
&=&\tr\left[{1\over N}\hT_\beta^\dag\sum_\alpha{\cal
C}_\epsilon(\alpha)e^{(2\pi \rmi /N)\alpha\wedge\beta'}\hT_\beta'\right]\nonumber\\
&=&{1\over N} \tr[\widetilde{\cal C}_\beta'\hT_\beta^\dag\hT_\beta']=\widetilde{\cal
C}_\epsilon(\beta)
\end{eqnarray}
where $\widetilde{\cal C}_\epsilon\beta)$ is the DFT of ${\cal C}_\epsilon(\alpha)$ and is also a Gaussian
of complementary width $\propto 1/\epsilon$.
As a consequence if the chord function of a state $\hrho$ is say $\varrho(\alpha)$, the action
of $\De$ in the chrod representation is
\begin{equation}
	\label{eq:chordop}
\varrho'(\beta)=\widetilde{{\cal C}}_\epsilon(\beta)\varrho(\beta)
\end{equation}
and the long chords ($\beta\gg 1/\epsilon$) are supressed. 
The chord function is related to the DWF by a Fourier transform so the long chords are related
to the small structures in the wigner function. 
The small scale structures in the DWF are associated to quantum interference and
coherence. Therefore erasing the long chords is equivalent to erasing short wavelength
oscilations in the DWF resulting in a decoherent process.

In analogy with \equa{eq:chordop} the chord re[presentation provides a picture of what
the effect of the noise when composed with a unitary process.
Composition with a unitary map $U\odot U^\dag$ yields again a RUP 
(and thus a contractive map).
Let $\s=\De\circ\SOp{U}$, with $\SOp{U}=\hU\odot\hU^{\dag}$. Then in the chord representation
it is just
 \begin{equation}
	\label{eq:noisechord}
{\cal S}(\alpha\beta)=\widetilde{{\cal C}}_\epsilon(\alpha){\cal U}(\alpha\beta)
\end{equation}
where ${\cal U}(\alpha\beta)=\tr[\hT_{\alpha}^\dag \hU\hT_{\beta}\hU^\dag]$, and 
Therefore
the Gaussian function $\widetilde{{\cal C}}_\epsilon(\alpha)$ 
{\em modulates\/} the chord funtion of $\SOp{U}$ supressing the long chord components.
This is in fact a coarse graining of $\SOp{U}$ and we shall call $\s_\epsilon$ 
{\em coarse-grained
quantum propagator\/}.
\section{Chaos and Correspondence}
	\label{sec:ch-corr}
Let us first give the now standard definition of quantum chaos. By quantum chaotic systems we mean
those whose classical counterpart is chaotic. Usual approaches to understanding chaos and quantum mechanics range from the
concept of universality in random matrix theory \cite{bohigas}, and also since the
celebrated Gutzwiller\cite{gutz} trace formula the use of periodic orbits in the
computation of the eigenvalues and eigenfunctions\cite{vergini,heller} became fundamental.

Recently yet another different approach has been explored. For strongly chaotic
(Axiom-A\footnote{Axiom-A or Anosov-A are uniformly hyperbolic systems with strong
mixing.})
systems correlation functions, as well as any time dependent quantity decay exponentially.
The decay rates are closely related to the spectrum of the propagator of densities phase space. 
The idea is to establish, be it numerically\cite{garma,garma2,braun,manderfeld}, 
experimentally\cite{sridhar,jalabert} 
or theoretically\cite{nonn}, a relationship between the classical decay rates and the spectrum of
the noisy quantum propagator.
\subsection{Ruelle-Pollicot Resonances}
	\label{sec:RP}
Let $M$ be an invertible area preserving map.

Formally the
evolution of a distribution $\rho(\nx{x})$ in phase space is given by the
Liouville equation
\begin{equation}
\frac{\pd\rho}{\pd t}={\cal L}\rho=\{H,\rho\}
\end{equation}
where the Liouville operator ${\cal L}$ is the generator of the Perron Frobenius (PF) 
operator ${\cal P}$
\begin{equation}
\rho(\nx{x},t)={\cal P}_t\rho(\nx{x},0).
\end{equation}
When possible we can reduce the dynamics to a stroboscopic area preserving map $M$ and
the solution in terms of the PF and $M$ can be ritten as\cite{chaosbook} 
\begin{equation}
	\label{eq:PF}
\rho_{n+1}={\cal P}\rho_n=
\int{\rm d}\nx{x}'\delta(\nx{x}-M(\nx{x}'))\rho_n(\nx{x}').
\end{equation}
The PF is unitary in the space of square integrable functions $\Ldos$. 

Effective irreversibility in chaotic systems manifests in the exponential 
decays of correlation
functions. For two functions $f,g$ in phase space we will use the following simplified
expresion for the correlation function
\begin{equation}
	\label{eq:correl}
C_t(f,g)=(f,{\cal P}^tg)-(I,f)(I,g)
\end{equation}
where  $(\ , \ )$ denotes inner product and $I$ is the uniform density. 
For strongly mixing systems $C_t(f,g)$ goes to zero as
$t\to\infty$.
The decay rates  are the so-called Ruelle-Pollicott(RP) resonances.
For Anosov A systems they are the poles of 
the resolvent of the PF in a higher Riemannian sheet\cite{ruel}. 
There are different methods to unveil the resonances from the PF.
One method is directly interpolating the exponential decay 
in the time evolution\cite{baladi}. Other more
sophisticated consist in Fredholm determinant techniques\cite{christ} and there are also cycle
expansion\cite{gaspard} and trace formula approaches\cite{braun}. Markov Partitions were
introduced in the study of decay of correlations by Brini \etal\cite{brini}

\begin{figure}
\begin{center}
\includegraphics*[215,343][382,495]{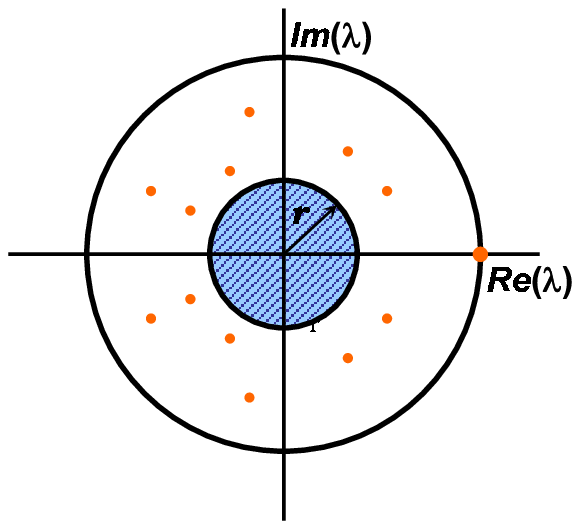}
\caption{Schematic representation of the spectrum of ${\cal P\/}$. The dots represent the point spectrum
($\sigma_{\rm ps}$) while the circle of radius $r$ represents the essential
spectrum.\label{fig:esquema}}
\end{center}
\end{figure}

The RP resonances can also be related to the
spectrum of ${\cal P}$ acting on a functional space different from $\Ldos$. 
Recently Blank \etal\cite{blank} provided an accurate theoretical 
description of the spectrum of ${\cal P}$ for
Anosov A systems.
The idea is that
for area preserving maps if ${\cal P}$ is restricted to spaces like $\mathbb{L}^p$ 
where the
norm essencially provides information about size, the spectrum does not provide much 
information
about the dynamics. Therefore they use
strongly anisotropic Banach spaces of functions especially 
adapted to the dynamics\footnote{Similar spaces were also used by Rugh\cite{rugh}.} 
composed of asymmetrical sets of
functions which are smooth along the unstable direction and increasingly 
singular along the stable
direction. They showed the operator restricted to this space to be {\em quasi-compact\/}. 
Moreover they showed that the spectrum is stable under pertubation and proposed a
discretization scheme to compute the spectrum from finite size matrices. 
The spectrum thus found (represented sechematically on \fig{fig:esquema}) 
can be described as follows: a point spectrum, with finite
multiplicity
\begin{equation}
\sigma_{\rm ps}=\{\lambda_i, r<\lambda_i<\lambda_0=1\}
\end{equation}
for some radius $r>0$, and an essential spectrum inside the circle of radius $r$. Notice that for
hyperbolic systems the gap between $\lambda_0=1$ and $\lambda_1$ is strictly non-zero. This is essential
for exponential decays to occur. 
Now assuming that the eigenvalues are nondegenerate (and that $r\ll 1$)
then we can make an approximate spectral decomposition of ${\cal P}$ as
\begin{equation}
	\label{eq:spexpans}
{\cal P}\approx \sum_i \lambda_i {\cal R}_i({\cal L}_i,\ .\ )
\end{equation}
where ${\cal R}_i$, (${\cal L}_i$) are the right (left) eigenfunctions.
Using this spectral decomposition explicitly in \equa{eq:correl} we get an 
expansion of $C_{fg}$ like
\begin{equation}
	\label{eq:corel-spec}
C_{fg}(t)= \sum_i \lambda_i^t(f,{\cal R}_i)({\cal L}_i,g)
\end{equation}
clearly dominated by the point spectrum $\sigma_{\rm ps}$. 
From \equa{eq:corel-spec} it is clear that in the limit $t\to \infty$ we have
$C_{fg}(t)\sim \lambda_1^t(f,{\cal R}_1)({\cal L}_1,g)$, and the importance of the largest
(in modulus)
RP resonance becomes evident.
The preceding analysis allows us to identify the complex eigenvalues 
$\lambda_i$ in the point spectrum of $\sigma_{\rm sp}$ of ${\cal P}$ with the RP resonances.

There are many methods (in use lately) to compute the classical PF spectrum. We briefly
describe two of the most popular.
The first one is a
truncation method used for example by Manderfeld\etal\cite{manderfeld}, 
Khodas \etal\cite{khodas}, Hasegawa
and Saphir\cite{hase} and was nicely summerized by Fishman\cite{fishman} in
the following way:
\begin{enumerate}
\item Introduce a basis ordered by increased resolution. Typically Fourier
modes, Legendre
polynomials\cite{hase} or
spherical harmonics\cite{manderfeld}.
\item Truncate to a desired dimension $N$ leaving the rapidly oscillating components out. 
This acts as coarse graining of ${\cal P}$.
\item Compute the eigenvalues of the truncated operator and as $N$ is increased
observe that there is a part that {\em freezes\/}  and is independent of $N$.
\item Compare the frozen eigenvalues with the resonances obtained by analytic continuation of
the resolvent.
\end{enumerate}
It should be duly noted that this method was successfully implemented also for systems with a mixed
phase space\cite{weber}.

The second method consists of implementing a {\em coarse graining\/}\cite{nonn,agam}
of ${\cal P\/}$ by convoluting it with a Gaussian kernel of width $\epsilon$
\begin{equation}
{\cal P}_\epsilon\stackrel{{\rm def}}{=}{\cal D}_\epsilon\circ{\cal P\/}.
\end{equation}
Essentially this is achieved by just replacing the delta functions in \equa{eq:PF} by normalized
(and periodized in the case of the torus) Gaussians of width $\epsilon$.
This  kind of coarse graining is equivalent to adding diffusive noise in the equations of
motion.\cite{gaspard2}
In order to perform the calculations a discretization of phase space is needed. The RP
resonances are the eigenvalues of the coarse grained PF, ${\cal P}_\epsilon$, after doing an
extrapolation of $N\to\infty$ and $\epsilon\to 0$.\footnote{The limits are non-commutative,
\ie if $\epsilon\to 0$ faster than $N\to\infty$ then the eigenvalues do not ``freeze'' and the
spectrum becomes unitary again.\cite{nonn} However no general, and optimal, relation
$\epsilon(N)$ has been found.}
\subsection{Spectrum of the Noisy Quantum Propagator} 
In a recent work Nonnenmacher\cite{nonn}
proved the following theorem (which we state in a simplified form):

\smallskip
\noindent
\textsf{\textbf{Theorem}} (Nonnenmacher\cite{nonn}).
\textit{
For any smooth map on the torus and any fixed coarse graining parameter $\epsilon>0$, the
spectrum of the coarse grained propagator $\s_\epsilon=\SOp{D}_\epsilon\circ\SOp{U}$,
converges in the classical limit $N\to \infty$ to the point spectrum of the classical coarse-grained
propagator ${\cal P}_\epsilon={\cal D}_\epsilon\circ{\cal P}$.}

\noindent
Where both coarse grained propagators are like the ones described in the previos section and
the noise is assumed to be Gaussian with characteristic width $\epsilon$.
\smallskip

Thus the introduction of noise in the quantum propagator makes the classical decay rates
(RP resonances) emerge naturally.
The problem which persists is the need to explore the large $N$ region because then the
propagator of density matrices is a, much larger, $N^2\times N^2$
matrix.
However the decays will be ruled by the largest eigenvalues (in modulus) so in fact only a
small group of the leading
of the eigenvalues is really relevant.
In what follows we describe some methods used recently to compute the relevant part of
the spectrum.
\section{Numerics}
	\label{sec:num}
The goal is to explore regions of $N$ ``large''. Thus the
matrix to diagonalize, in principle, is $N^2\times N^2$. 
In the large $N$ limit this may represent a problem in terms of 
time and memory requirements. However, as was previously pointed out, only the largest
eigenvalues (in modulus) play a significant role in the long time bahavior of
observables. Therefore approximation methods can be used to significantly reduce the
eigenvalue problem.

In what follows we describe three different methods to compute the leading spectrum. The
first two are described briefly. They have the advantage that the only limitation on
the number of eigenvalues to compute is the size of the matrices to digonalize. 
The third method, described in more detail, has the advantage that the first few
(typically 10) eigenvalues can be computed with very small matrices (typically
$10\times 10$).
\subsection{Truncation Method:}
Again, the truncation method described in \sect{sec:RP} was succesfully used by Manderfeld
\etal to obtain a nonunitary, coarse grained quantum propagator and to show that classical and
quantum evolutions look alike (in the $N\to \infty$ limit) and that decay rates can be computed
from the truncated propagator. However, this approach can have problems of interpretation if
not of physical realizability. Spina and Saraceno\cite{spina} recently showed that when the
quantum propagator is truncated in the way described, 
which is essentially a sharp truncation,
the whole operation, unitary evolution followed by noise is not
completely posititve and cannont be obtained as a quantum 
operation.\footnote{There can be some
debate\cite{debate} on weather complete positivity is an essential requirement. We favour the
opinion that it is a necesary requirement.}
\subsection{Chord Function Method:}
On the other hand using 
the diffusive noise of \sect{sec:diff}  and the chord representation a ``smooth'' 
truncation can be implemented (see Spina and Saraceno\cite{spina} and
Aolita, \etal\cite{aolita}). Indeed the complete positivity is directly related to the
positivity of the coeficients ${\cal C}_\epsilon(\alpha)$. Now since  
${\cal C}_\epsilon(\alpha)$ is a (periodic) Gaussian then the spectrum $ \widetilde{\cal
C}_\epsilon(\beta)$ is also a Gaussian of complementary width (proportional to $1/\epsilon$).
Therefore a large part of the elements of $\s_\epsilon$ in the chord representation will be
very small. If one sets for example the elements such that $\beta$ is larger than $1/(2\pi N)$
by some factor (bigger than one), exactly equal to zero, then, even though there will be
oscillations in  ${\cal C}_\epsilon(\alpha)$ (because it is the inverse DFT) they can be 
made negligibly small just by moving the truncation parameter. Using this method the matrices
to diagonalize can be reduced to a dimension of order 
$N\times N$ and the precision obtained for the resonances is
very good. The compromise lies in that using very small matrices to get the RP
resonances means using very strong noise. So $N\times N$ is kind of a lower bound in the size of
the eigenvalue problem to solve using the chord function method. 
\subsection{Iteration Method}
\begin{figure}[htb!]
\begin{center}
\scalebox{0.85}{
\includegraphics*[111,182][494,609]{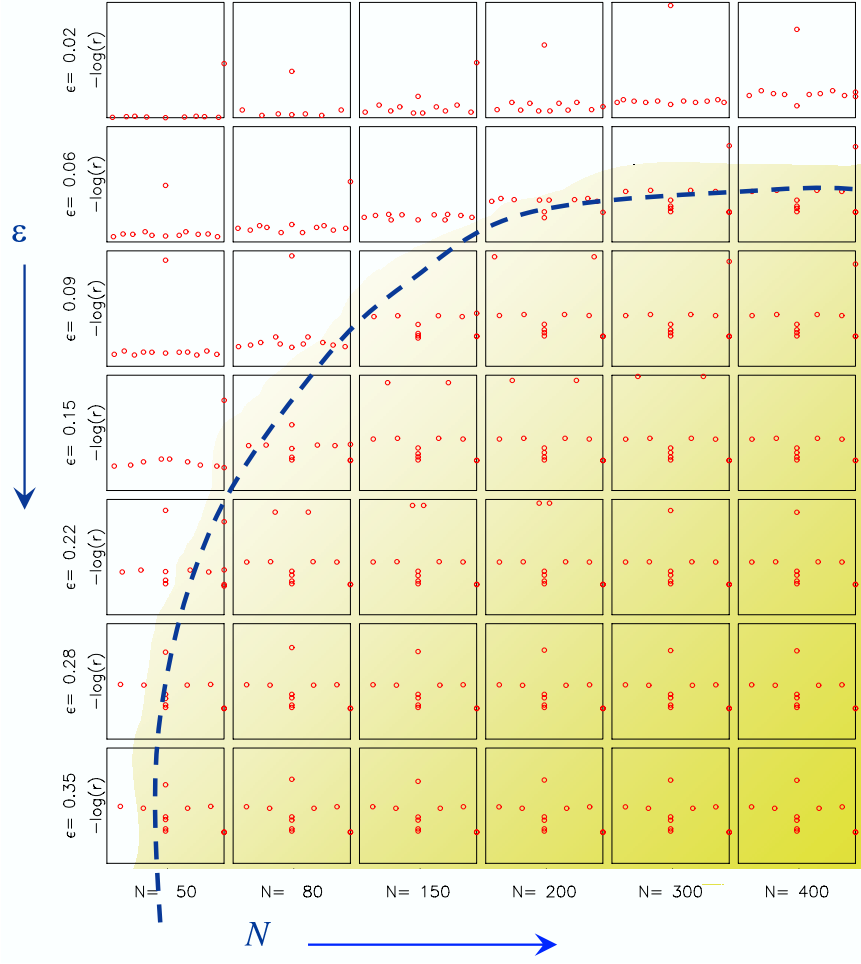}
}
\caption{ Spectrum of the propagator for the perturbed cat map ($k=0.01$) for 
different values of $\epsilon$ and $N$. 
The dashed line shows approximately the division in the space of the parameters 
between quantum (almost unitary) and noisy (almost classical) behavior.\label{fig:spect}}
\end{center}
\end{figure}

The iteration method is a kind of variation of some known iteration
methods\cite{agam,florido} ({\em e. g.\/} Lanczos iteration method\cite{golub}). The main
requirement is that the spectrum is of the form 
$1>|\lambda_1|\geq|\lambda_2|\geq\ldots\geq|\lambda_k|\geq\ldots$. The
convergence  depends strongly on the size of the gap $(1-|\lambda_1|)$.
Consider the two sets of iterated states 
\begin{equation}
{\cal B}_f=\{\hrho_0,\hrho_1,\ldots,\hrho_k,\ldots\};\ \ 
{\cal B}_b=\{\hrho_0,\hrho_{-1},\ldots,\hrho_{-k},\ldots\}
\end{equation}
where $\hrho_0$ is an arbitrary initial state and 
\begin{equation}
\hrho_k=\s_\epsilon^k\hrho_0;\ \ 
\hrho_{-k}=\s_\epsilon^{\dag\,k}\hrho_0
\end{equation}
Then we write a kind of projection of $\s_\epsilon$ onto the subspace spanned by
this (non-orthogonal) sets
\begin{equation}
\big[\s_\epsilon\big]_{ij}=\tr(\hrho_{-i}^\dag\s_\epsilon\hrho_j)
\end{equation}
and solve the generalized eigenvalue equation
\begin{equation}
	\label{eq:det}
\text{Det}\left[\big[\s_\epsilon\big]_{ij}-
\lambda\big[\SOp{O}\big]_{ij}\right]=0
\end{equation}
where $[\SOp{O}]_{ij}=\tr[\hrho_{-i}^\dag\hrho_j]$ is the matrix of overlaps.
Now if the spectrum has the required form, then both $[\s_\epsilon]_{ij}$
and $[\SOp{O}]_{ij}$ decay very rapidly with $i,j$ so depending on the
precision required a truncation size $k$ can be chosen. Typicaly in all the
computations made\cite{garma,garma2}, for $N$ between 50 and 450, the matrices
diagonalized where of sizes from $10\times 10$ to $15\times 15$.
In \fig{fig:spect} some results obtained\cite{garma,garma2} with this method are displayed. The shaded region
indicates the region in the prameters $N$ and $\epsilon$ inside which the spectrum stabilizes
and can be considered classical. 
Care should be taken when approaching the limits $N\to\infty$ and $\epsilon\to 0$ because
this limits do not commute. If $\epsilon$ goes to zero faster than $N$ to infinity then
unitarity is recovered.
There seems to be a kind of optimal function
$\epsilon(N)$ to approach the classical limit. The fit is not easy to compute because 
it is very much system dependent.

In order to show that the values computed are the desired first $k$ largest
eigenvalues we assume, for simplicity, that the spectrum is non-degenerate. The
propagator can be written in its spectral decomposition form as
\begin{equation}
\s_\epsilon\hrho=\sum_{i}\lambda_i\hat{R}_i\tr[\hat{L}_i^\dag\hrho]
\end{equation}
where $\hat{R}_i$ ( $\hat{L}_i$) is the right (left) eigenfunction. We can
assume also that the overlap between the initial state $\rho_0$ and the right
and left eigenfunctions is non-zero. Then \equa{eq:det} can be re-written as
\begin{equation}
\text{Det}[
\grkbf{\cal K}^T(\s_\epsilon,\hrho_0,k)\s_\epsilon\grkbf{\cal K}(\s_\epsilon,\hrho_0,k)
-\lambda\grkbf{\cal K}^T(\s_\epsilon,\hrho_0,k)\grkbf{\cal K}(\s_\epsilon,\hrho_0,k)
]=0
\end{equation}
where $\grkbf{\cal K}(\s_\epsilon,\hrho_0,k)$ is the Krylov matrix whose columns
are the succesive iterates of the initial state $\hrho_0$ up to order $k$ (which
is the truncation size we choose)
\begin{equation}
\grkbf{\cal K}(\s_\epsilon,\hrho_0,k)=[\hrho_0,\s_\epsilon,\hrho_0,\ldots,
\s_\epsilon^{k-1}\hrho_0]
\end{equation}
and $T$ denotes matrix transposition.
Then using the expansion of $\hrho_0$ both in terms of $\hat{R}_i$ and
$\hat{L}_i$ it can be shown\cite{garma2} that the determinant can be written
as
\begin{equation}
	\label{eq:det2}
\text{Det}[\grkbf{\Lambda}^\dag\grkbf{\Xi}\grkbf{\Lambda}]=0
\end{equation}
with
\begin{equation}
\grkbf{\Lambda}=\left(
\begin{array}{ccccc}
1&\lambda_0&\lambda_0^2&\cdots&\lambda_0^{k-1}\\
1&\lambda_1&\lambda_1^2&\cdots&\lambda_1^{k-1}\\
\vdots   &\vdots   &\vdots  &\ddots& \vdots \\
1&\lambda_{k-1}&\lambda_{k-1}^2&\cdots&\lambda_{k-1}^{k-1}
\end{array}
\right)
\end{equation}
and
\begin{equation}
\grkbf{\Xi}= \left(
\begin{array}{cccc}
\alpha_{0}\beta_{0}(\lambda_0-z)&0&\cdots&0\\
0&\alpha_{1}\beta_{1}(\lambda_1-z)&\cdots&0\\
\vdots&\vdots&\ddots&\vdots\\
0&\cdots&0&\alpha_{k-1}\beta_{k-1}(\lambda_{k-1}-z)
\end{array}
\right).
\end{equation}
Now $\grkbf{\Lambda}$ is a Vandermonde matrix and from
m the asumption of non-degeneracy, the determinant is non-zero. 
Thus the solution of \equa{eq:det2} is just given
by the zeros of the determinant of $\grkbf{\Xi}$ which 
are exactly the desired first $k$
eigenvalues.
Even though we do not discuss precision we can say that the efficiency of the
method depends mainly on two things. First, there has to be a finite gap between
1 and $\lambda_1$. The second, and very important also, is that the map
should be easy to implement without constructing the whole $N^2
\times N^2$ superoperator.
\subsection{Long time behavior}
	\label{sec:LT}
Exponential stretching (with the rate given by the Lyapunov
exponent) is the cause that correspondence is lost much faster than for
regular systems. This discrepancy is observed in phase space quasi-probability  
distributions such as the Wigner
function\cite{dittrich,kolovsky} in the form of sub-Planck structure.\cite{zurekRMP} 
For isolated systems, the characteristic time for the loss of
correspondence in the Wigner function
is usually called Eherenfest\cite{berman} time. It is generally a very short time and can be
expressed as $\tau_E=\lambda^{-1}\ln(S/\hbar)$, where $\lambda$ 
is the largest Lyapunov exponent and $S$ is
some characteristic action.

Zurek and Paz conjectured\cite{zurek} that loss of correspondence in chaotic
systems would be recovered by introducing decoherence. Moreover they stated
that initially the linear entropy 
\begin{equation}
S_n=-\ln\left[\tr[\hrho_n^2]\right],
\end{equation}
(which is the natural logarithm of the purity) should grow linearly with $n$ (\ie time), 
with a slope given by the largest lyapunov exponent of the corresponding classical
system. This lyapunov behavior would have to be independent of the coupling
strengh, that is the strengh of the noise, for a large range of values of the
coupling. This statement was verified numerically\cite{bianucci,garma}. The
lyapunov growth takes place until the Eherenfest time. Then the linear entropy
saturates asymptotically to the value $\ln(N)$ that corresponds to the maximally
mixed state $\hrho_\infty=\II/N$. 

Another quantity of interest with similar short time behavior is the Loschmidt echo. It was
proposed\cite{peres} as a
measure of irreversibility and could be used to characterize quantum chaos. Many recent 
works\cite{jalabert,losch} show that
the lyapunov regime indeed exists for classically chaotic systems. 
This quantity gained interes recently
in the context of quantum information because of its close relation to fidelity. 

There is yet another classical regime that
can be identified. If we assume that there
are nondegeneracies then in analogy \equa{eq:spexpans} a spectral decomposition can be
made but for the quantum noisy propagator
\begin{equation}
\s_\epsilon\hrho=\sum_{i=0}^{N^2}\lambda_i \tr\left[\hL_i^\dag\,\hrho\right]\,\hR_i
\end{equation}
where $\hR_i$ and $\hL_i$ are the right and left eigenoperators of $\s_\epsilon$,
and the eigenvalues are ordered by decreasing modulus (starting with
$\lambda_0=1$). 
Then, for example, the autocorrelation function behaves asymptotically as
\begin{eqnarray}
C_n&=&\tr\left[\hrho_0\hrho_n\right]-\tr\left[{\II\over N}\hrho_0\right]^2\nonumber \\
  &=&\lambda_0\,\tr\left[\hrho_0\hrho_\infty\right]\tr\left[\hrho_\infty\hrho_0\right]\nonumber 
  +\sum_{i=1}^{N^2}\lambda_i^n\tr\left[\hrho_0\hR_i\right]\tr\left[\hL_i^\dag\hrho_0\right]
  -{1\over N^2}\nonumber \\
 &\sim&\lambda_1^n\,\tr\left[\hrho_0\hR_1\right]\tr\left[\hL_1^\dag\hrho\right].
\end{eqnarray}
where $\lambda_0=1$ and $\hrho_\infty$ as defined in \equa{eq:uniform}.
In Table \ref{tabla} the long time behavior 
of the autocorrelation function, the linear entropy 
(logarithm of the purity)
and the Loschmidt echo is summerized. In all cases the logarithm grows linearly with time ($n$) and the
slope is determined by $\lambda_1$. Now taking $N$ and $\epsilon$ in the correct region of
\fig{fig:spect}, $\lambda_1$ is identified with the largest RP resonance of the classical system.
Thus another classical property has emerged from the quantum system by the addition of noise.
\begin{table}
\begin{center}
\begin{tabular}{|c|c|l|l|}
\hline
\textsf{\textbf{Quantity}}&\textsf{\textbf{Definition}}
			&\textsf{\textbf{Aymptotic behavior}}
			&\quad\textsf{\textbf{Ln}}\\
\hline\hline		
Autocorrelation & $\tr[\hrho_0\hrho_n]-1/N$ & \quad$\sim\ \lambda_1^n$
	& \quad$\sim\ n\lambda_1$ \\
\hline
Linear Entropy	& $-\ln[\tr[\hrho_n^2]]$ & \quad$\sim\ -2 n \ln[\lambda_1]$& 
	\quad$\sim\ -2 n \ln[\lambda_1]$\\
\hline
Loschmidt  echo & $\tr[\hrho_n\hrho'_n]$ & \quad$\sim\ \lambda_1^n(\lambda'_1)^n$&
	\quad$\sim\ n(\lambda_1+\lambda_1')$\\
\hline
\end{tabular}
\end{center}
\caption{Long time ehavior of the aurotcorrelation function$C_n$, the linear entropy $S_n$ 
and the Loschmidt echo $M_n$. In the case of $S_n$ and $M_n$ we measure the {\em approach\/} to
equilibrium and for tah purpose we explicitly subtracted the invariant state $\rho_\infty$ in order to
avoid saturation.\label{tabla}}
\end{table}
\begin{figure}[htb!]
\begin{center}
\includegraphics*[width=9.0cm]{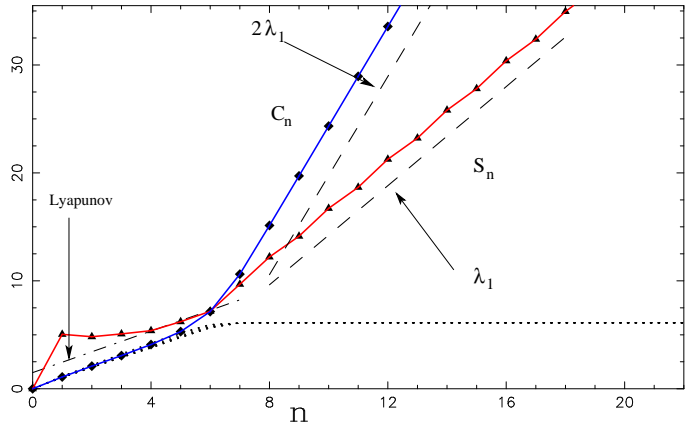}	
\caption{Time evolution of the linear entropy $S_n$ (diamonds) and the
autocorrelation function (triangles). The points result from the average of ten
initial conditions corresponding to coherent states centered at ten different
points. The Dot-Dash-Dot line indicates the initial Lyapunov behavior and the
Dashed lines indicate the RP behavior. The slopes of the RP lines were obtained
from the leading eignvalue computed using the iteration method. As it was
expected, the slope for
$S_N$ is exactly two times the slope of $C_n$. The dotted line shows the behavior if the uniform density
is not subtracted. \label{fig:timeplot}}
\end{center}
\end{figure}

The long time linear regime can be best observed in \fig{fig:timeplot} where $C_n$ and $S_n$ are plotted.
In order to uncover the long time regime for $S_n$ we subtracted $\rho_\infty$ from the initial states.
The initial Lyapunov regime is very well observed for $S_n$ and for longer times the RP regime is also
well defined. The fact that $C_n$ has no Lyapunov regime is well understood. For a chaotic system the
overlap between the initial state and the succesive iterates is a highly fluctuating function until after
the Eherenfest time (in which the phase space distribution has reached the edges of phase space). This is
the reason why we averaged over a number of initial states in order to get a smoother function.
The time graphs for the Loschmidt echo are very similar\footnote{See for example Fig 4 in
Garc\'{\i}a-Mata, \etal\cite{garma} and Fig 8(b) in Garc\'{\i}a-Mata and Saraceno\cite{garma2}}

\section{Concluding remarks}
Using some new numerical methods we established quantum classical 
correspondence for chaotic quantum maps 
throught the spectrum of the coarse-grained propagator. 
For a wide range of parameters, the spectrum could be related 
to the classical Ruelle Pollicott resonances.  
Many techniques to compute the spectrum where presented and some of the advantages of the iteration method (mainly for chaotic systems) could be appreaciated.
As an open question there remains the issue of establishing an 
optimal relation betwwen the noise strength $\epsilon$ and $\hbar$ ($N$) 
so that the classical limit can be attainedcorrectly. Recent work\cite{toscano} 
provides an expression such relation in the context of correspondence-breaking time. Although it looks promising we have no conclusive proof of its relation to the spectrum.   
\section*{Acknowledgments}
The authors profited from discusions with St\'ephane Nonnemacher. 
I. G.-M. thanks the hospitality at the {\em Max-Planck-Institute f\"ur Physik
komplexer Systeme\/} (Dresden), and the {\em Center for Nonlinear and Complex Systems\/} 
(Como) were part of this work was completed. I. G.-M. would also like to thank
thank Prof. Andreas Buchleitner, Andr\'e Ribeiro de 
Carvalho and Gabriel Carlo for enlightening discussions.
Financial support was provided by CONICET and ANPCyT.

\end{document}